\documentclass[12pt]{article}
\usepackage{epsf}
\usepackage{amsmath}
\usepackage{graphics}
\usepackage{cite}

\renewcommand{\thefootnote}{\#\arabic{footnote}}
\begin{document}

\newcommand{\gtrsim}{ \mathop{}_{\textstyle \sim}^{\textstyle >} }
\newcommand{\lesssim}{ \mathop{}_{\textstyle \sim}^{\textstyle <} }

\newcommand{\rem}[1]{{\bf #1}}

\renewcommand{\thefootnote}{\fnsymbol{footnote}}
\setcounter{footnote}{0}
\begin{titlepage}

\def\thefootnote{\fnsymbol{footnote}}

\begin{center}
\hfill hep-ph/0608039\\
\hfill August 2006\\
\vskip .5in
\bigskip
\bigskip
{\Large \bf Darwin and Phenomenology Beyond the Standard Model
\footnote{Talk at the Second World Summit on Physics,
San Cristobal, Galapagos, Ecuador, June 22-25, 2006}}

\vskip .45in

{\bf Paul H. Frampton}

{\em University of North Carolina, Chapel Hill, NC 27599-3255, USA}

\end{center}

\large

\vskip .4in
\begin{abstract}
After preamble about Darwin, my
talk described the conformality approach to extending the standard model
of particle phenomenology using an assumption of no conformal
anomaly at high energy. Topics included quiver gauge theory, the
conformality approach to phenomenology, strong-electroweak
unification at 4 TeV, cancellation of quadratic divergences,
cancellation of U(1) anomalies, and a dark matter candidate.
\end{abstract}
\end{titlepage}

\renewcommand{\thepage}{\arabic{page}}
\setcounter{page}{1}
\renewcommand{\thefootnote}{\#\arabic{footnote}}

\newpage

\large

\noindent {\it Preamble}

\bigskip

I would first like to thank the organizers, Carlos Montufar
and Bruce Hoeneisen for the invitation.
Since this is the only evening talk of the meeting and conferees
are still coming in from dinner, I have a few minutes to talk 
about the local hero, Charles Darwin.

Of the participants at this conference, I was born nearest to Shrewsbury
which is Darwin's birthplace. I was born and raised in Worcestershire
from which Shrewsbury is just across the county line in Shropshire. When I was
growing up the answer to the question ``Who was the greatest scientist
ever born in England?" was unambiguously ``Newton". In the fifty years
since then that has changed and Darwin is now
placed on a pedestal equally as high as Newton's. 
Darwin's fame is helped by results of polls in America showing half 
of the people do not believe in evolution.

Indeed, when I was growing up, Shrewsbury had no museum or even a marker
on Darwin and most people probably did not know of him. Now in
Shrewsbury there is a large statue of a seated elderly Darwin, even
larger than the one of a standing young Darwin here on Frigatebird Hill.
There is also a Darwin museum in Shrewsbury and it is now an important
visit in the well-organized English tourist industry.

One lecturer on the first day made an analogy between Darwin's visit
to the Galapagos in 1835, towards the end of his five year voyage on
the Beagle, and the commissioning of the LHC next year. Let me pursue
his analogy further. Firstly, after Darwin returned to England it
took him 4 or 5 years to figure out the theory of evolution: there
was no sudden epiphany here in the Galapagos. It may take a similar time
to deduce the correct theory from LHC data.
Secondly, although the Turtles, Iguanas and Boobies are some of
the more obvious
inhabitants of the Galapagos the best examples for Darwin's theory
were the beaks of the small finches which could not fly
in between the islands. Perhaps one should look for
the analogs of the beaks of the finches which would be small details in 
a less conspicuous subset of the LHC data.

It is startling that these two greatest scientists, Newton and Darwin,
were both born in England with one per cent of the world's population. 

\noindent I have boasted enough about England. Now I can begin my talk.

In case there is any misunderstanding, this is not a public lecture
as someone asked me this morning.
What the organizers requested was that speakers ``bring original
ideas on viable extensions of the standard model". What I will 
talk about is certainly 
original; whether it is viable or not 
may have to await the LHC or at least the discussion
section but I will do my best.

\bigskip
\bigskip
\bigskip

\noindent {\it Introduction}

\bigskip

In this talk I will describe an approach to extending the standard model
of particle phenomenology to make predictions for the Large Hadron Collider
which is based on conformality, defined as the absence of conformal
anomaly at high energy.

\bigskip

The contents of the talk are as follows:

\bigskip

\noindent 1. Quiver gauge theories.

\noindent 2. Conformality phenomenology.

\noindent 3. 4 TeV Unification.

\noindent 4. Quadratic divergences.

\noindent 5. Anomaly Cancellation and Conformal U(1)s

\noindent 6. Dark Matter Candidate.

\noindent 7. Summary.

\bigskip
\bigskip

Before I start here is a concise history of string theory:

\bigskip

\begin{itemize}

\item Began 1968 with Veneziano model.

\item 1968-1974. Dual resonance models for strong interactions.
Replaced by QCD around 1973. DRM book 1974. Hiatus 1974-1984

\item 1984. Cancellation of hexagon anomaly.

\item 1985. $E(8) \times E(8)$ heterotic strong compactified on
Calabi-Yau manifold gives temporary optimism of TOE.

\item 1985-1997. Discovery of branes, dualities, M theory.

\item 1997. Maldacena AdS/CFT correspondence relating 10 dimensional
superstring to 4 dimensional gauge field theory.

\item 1997-present. Insights into gauge field theory including possible
new states beyond standard model. String not only as quantum gravity but as powerful tool
in nongravitational physics.

\end{itemize}

\bigskip
\bigskip
\bigskip
\bigskip

QUESTION: What is meant by {\it duality} in theoretical physics?

\bigskip
\bigskip

ANSWER: Quite different looking descriptions of the same underlying theory.

\bigskip
\bigskip

The difference can be quite striking. For example, the AdS/CFT correspondence describes
duality between an ${\cal N}=4, d = 4$ SU(N) GFT and a $d = 10$ superstring.
Nevertheless, a few non-trivial checks have confirmed this duality.

\bigskip
\bigskip

In its simplest version, one takes a Type IIB
superstring (closed, chiral) in $d = 10$ and one compactifies on:

\[ 
~~~~~~~(AdS)_5 ~~~~~~~\times~~~~~~~ S^5
\] 

\bigskip
\bigskip

We recall old results on the perturbative finiteness of 
${\cal N}=4$ SUSY Yang-Mills theory:

\bigskip 

\begin{itemize}

\item Was proved by Mandelstam,
Nucl. Phys. B213, 149 (1983); P.Howe and K.Stelle, ICL preprint (1983), Phys.Lett. {\bf B137,} 135 (1984). 
L.Brink, talk at Johns Hopkins Workshop on Current Problems in High Energy Particle Theory, 
Bad Honnef, Germany (1983).

\bigskip

\item The Malcacena correspondence is primarily aimed at the
$N \rightarrow \infty$ limit
with the 't Hooft parameter of N times the
squared gauge coupling held fixed.

\bigskip

\item Conformal behavior assumed valid here also for finite N.

\bigskip

\item After exploiting initially the duality of gauge theory with string theory, we shall
therafter focus exclusively on the gauge theory description.

\end{itemize}

\newpage

\noindent SECTION 1. QUIVER GAUGE THEORIES

\bigskip

Breaking supersymmetries:

\bigskip

To approach the real world, one needs less or no supersymmetry in the (conformal?) gauge theory.

\bigskip

By factoring out a discrete (abelian) group and composing an orbifold:

\bigskip

\[ ~~~~~~S^5 / \Gamma ~~~~~~~\]

\bigskip

one may break ${\cal N} = 4$ supersymmetry to
${\cal N} = 2, ~~~~1,$ or $~~~0$. Of special interest is the ${\cal N} = 0$ case.

We may take $\Gamma = Z_p$ which identifies $p$ points in ${\cal C}_3$.

\bigskip

The rule for breaking the ${\cal N} = 4$ supersymmetry is:

\bigskip

\[ 
~~~~ \Gamma \subset SU(2)~~~~\Rightarrow~~{\cal N} = 2 
\]

\[ 
~~~~ \Gamma \subset SU(3)~~~~\Rightarrow~~{\cal N} = 1 
\]

\[ 
~~~~ \Gamma \not\subset SU(3)~~~~\Rightarrow~~{\cal N} = 0 
\]

\bigskip

In fact to specify the embedding of $\Gamma = Z_p$ we need to identify three integers $(a_1, a_2, a_3)$:

\[ 
~~ {\cal C}_3 :~~(X_1, X_2, X_3)~~\stackrel{Z_p}{\rightarrow}~(\alpha^{a_1} X_1, \alpha^{a_2} X_2, \alpha^{a_3}X_3)  
\]

with

\[ 
~~ \alpha = exp \left( \frac{2 \pi i}{p} \right)  
\]

\newpage

Matter representations:

\bigskip

\begin{itemize}

\item The $Z_p$ discrete group
identifies $p$ points in ${\cal C}_3$.

\bigskip

\item The N converging D3-branes
meet on all $p$ copies, giving a gauge group:
$U(N) \times U(N) \times ......\times U(N)$.

\bigskip

\item The matter (spin-1/2 and spin-0)
which survives is invariant
under a product of a
gauge transformation and a $Z_p$ transformation.

\end{itemize}

\bigskip

One can draw $p$ points and arrows for $a_1, a_2, a_3$.

\bigskip

\[ e.g.~~~~~Z_5~~(1, 3, 0) \]

\begin{figure}
\begin{center}
\epsfxsize=4.0in
\ \epsfbox{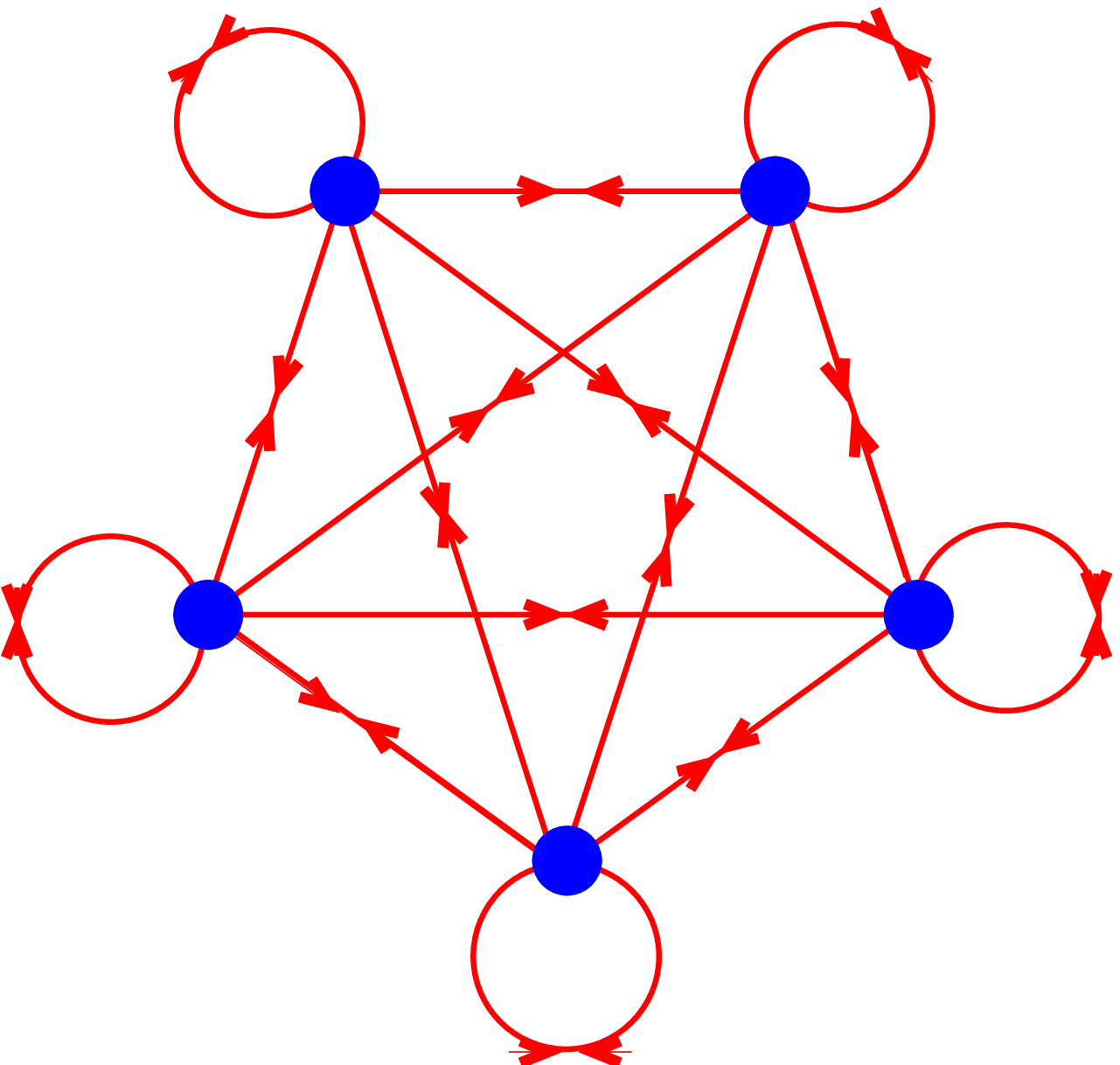}
\end{center}
\end{figure}

\noindent Quiver diagram    (Douglas-Moore).

\bigskip

\noindent Scalar representation is: $\sum_{k=1}^{3}\sum_{i = 1}^{p} (N_1, \bar{N}_{i \pm a_k})$ 

\newpage

For fermions, one must construct the {\bf 4} of R-parity $SU(4)$:

\bigskip

From the $a_k = (a_1, a_2, a_3)$ one constructs the 4-spinor $A_{\mu} = (A_1, A_2, A_3, A_4)$ :

\bigskip

\[ A_1 = \frac{1}{2} (a_1 + a_2 +a_3) \]

\[ A_2 = \frac{1}{2} (a_1 - a_2 -a_3) \]

\[ A_3 = \frac{1}{2} (- a_1 + a_2 - a_3) \]

\[ A_4 = \frac{1}{2} (- a_1 - a_2 +a_3) \]

\noindent These transform as $exp \left( \frac{2 \pi i}{p} A_{\mu} \right)$ and the invariants may again be derived (by a different diagram):

\begin{figure}
\begin{center}
\epsfxsize=4.0in
\ \epsfbox{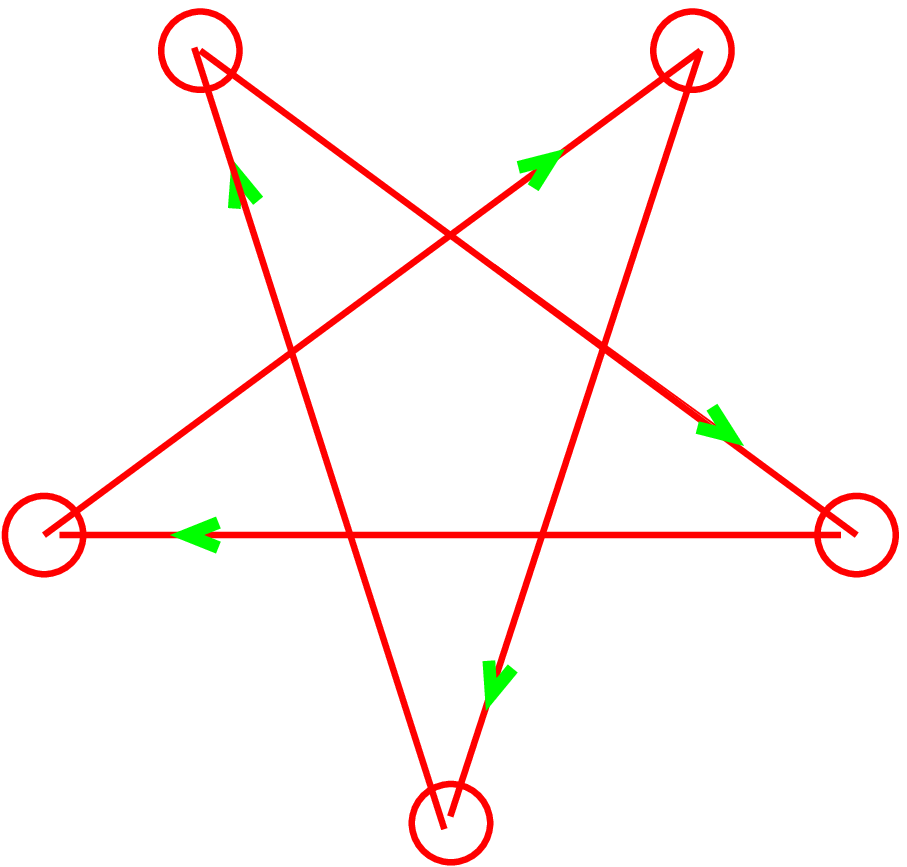}
\end{center}
\end{figure}

\[ ~~~e.g. A_{\mu} = 2;~~~p = 5. \]

\bigskip

\noindent These lines are oriented. Specifying the four $A_{\mu}$ is equivalent to the three $a_k$ and
group theoretically more fundamental

\bigskip

\noindent  One finds for the fermion representation

\bigskip

\[ \sum_{\mu = 1}^{4} \sum_{i = 1}^{p} ( N_i, \bar{N}_{i + A_{\mu}}) \]

\bigskip
\bigskip
\bigskip

\begin{center}

\bigskip

-----------------------

\bigskip

\end{center}

\noindent Two interesting reference for hierarchy and naturalness are:

\bigskip

\noindent K.G. Wilson, Phys. Rev. {\bf D3,} 1818 (1971) - already discussed in the late 1960s

\bigskip

\noindent K.G. Wilson hep-lat/0412043 -- what a difference 3 decades make.

\bigskip

\begin{center}

\bigskip

-----------------------

\bigskip

\end{center}

\noindent We also note that superconformal symmetry exemplified by ${\cal N} = 4$ SU(N) 
Yang-Mills in 1983 and related to string theory in 1997 for infinite N can be lessened to 

\bigskip
\bigskip

\noindent supersymmetry which answered, though as we shall see not uniquely,
Wilson's objection, rescinded in 2004,  about quadratic divergences in 1974 and generated
$> 10^4 $ papers.

\bigskip
\bigskip

\noindent or conformality according to  ${\cal N} = 4 \rightarrow {\cal N} = 0$  
by orbifolding $\rightarrow U(N)^n$ with, so far, $< 10^2$ papers.

\bigskip
\bigskip

\noindent Within a few years the Large Hadron Collider can experimentally discriminate between
the two possibilities.

\newpage

\noindent SECTION 2. CONFORMALITY PHENOMENOLOGY

\bigskip
\bigskip

\begin{itemize}

\item Hierarchy between GUT scale and weak scale
is 14 orders of magnitude. Why do these two very different
scales exist?

\bigskip

\item How is this hierarchy of scales stabilized
under quantum corrections?

\bigskip

\item Supersymmetry answers the second question but not the first.

\end{itemize}

\bigskip
\bigskip

\noindent The idea is to approach hierarchy problem by Conformality at TeV Scale.

\bigskip

\begin{itemize}

\item Will show idea is possible.

\bigskip

\item Explicit examples containing standard model states.

\bigskip

\item Conformality more rigid constraint than supersymmetry.

\bigskip

\item Predicts additional states at TeV scale for conformality.

\bigskip

\item Gauge coupling unification.

\bigskip

\item Naturalness: cancellation of quadratic divergences.

\bigskip

\item Anomaly cancellation: conformality of U(1) couplings.

\bigskip

\item Dark Matter candidate.

\end{itemize}

\newpage

\noindent Conformality as hierarchy solution.

\bigskip
\bigskip

\begin{itemize}

\item Quark and lepton masses, QCD and weak scales small compared to TeV scale.

\bigskip

\item May be put to zero suggesting:

\bigskip

\item Add degrees of freedom to yield GFT with conformal invariance.

\bigskip

\item 't Hooft naturalness since zero mass limit increases symmetry to conformal symmetry.

\end{itemize}

\bigskip
\bigskip

The theory is assumed to be given by the action:

\bigskip

\begin{equation}
S = S_0 + \int d^4x \alpha_i O_i
\end{equation}

\bigskip

\noindent where $S_0$ is the action for the conformal theory and the $O_i$ are operators
with dimension below four which break conformal invariance softly.

\bigskip

\noindent The mass parameters $\alpha_i$ have mass dimension $4-\Delta_i$ where
$\Delta_i$ is the dimension of $O_i$ at the
conformal point.

\bigskip

\noindent Let $M$ be the scale set by the parameters $\alpha_i$ and 
hence the scale at which conformal invariance is broken. Then for $E >> M$ the couplings 
will not run while they start running for $E < M$. 
To solve the hierarchy problem we assume $M$ is near the TeV scale.

\newpage

Experimental evidence for conformality:

\bigskip

Consider embedding the standard model gauge group according to:

\[ SU(3) \times SU(2) \times U(1) \subset \bigotimes_i U(Nd_i) \]

Each gauge group of the SM can lie entirely in a $SU(Nd_i)$
or in a diagonal subgroup of a number thereof.

\bigskip
\bigskip

\begin{center}

\bigskip

-----------------------

\bigskip

\end{center}

\noindent Only bifundamentals (including adjoints) are possible.
This implies no $(8,2)$, etc. A conformality restriction which is new and
satisfied in Nature!

\bigskip
\bigskip

\begin{center}

\bigskip

-----------------------

\bigskip

\end{center}

\noindent No $U(1)$ factor can be conformal and so hypercharge is quantized
through its incorporation in a non-abelian gauge group.
This is the ``conformality'' equivalent to the GUT charge quantization
condition in {\it e.g.} $SU(5)$!

\bigskip
\bigskip

\begin{center}

\bigskip

-----------------------

\bigskip

\end{center}

Beyond these general consistencies, there are predictions of
new particles
necessary to render the theory conformal.

\newpage

\noindent SECTION 3.~~~ 4 TeV UNIFICATION

\bigskip
\bigskip

\begin{itemize}

\item Above 4 TeV scale couplings will not run.

\bigskip

\item Couplings of 3-2-1 related, not equal, at conformality scale.

\bigskip

\item Embeddings in different numbers of the equal-coupling $U(N)$ groups lead to the 4 TeV scale
unification without logarithmic running over large desert.

\end{itemize}

\bigskip

When we arrive at $p = 7$ there are viable models. Actually three different quiver diagrams can give:

\bigskip

1) 3 chiral families.

2)Adequate scalars to spontaneously break $U(3)^7 \rightarrow SU(3) \times SU(2) \times U(1)$

and

3) ${\rm sin}^2 \theta_W = 3/13 = 0.231$

\bigskip

The embeddings of $\Gamma = Z_7$ in SU(4) are:

7A. ~~$(\alpha, \alpha, \alpha, \alpha^4)$

\bigskip

7B. ~~$(\alpha, \alpha, \alpha^2, \alpha^3)^*$ ~~ C-H-H-H-W-H-W

\bigskip

7C. ~~$(\alpha, \alpha^2, \alpha^2, \alpha^2)$

\bigskip

7D. ~~$(\alpha, \alpha^3, \alpha^5, \alpha^5)^*$ ~~ C-H-W-H-H-H-W

\bigskip

7E. ~~$(\alpha, \alpha^4, \alpha^4, \alpha^5)^*$ ~~ C-H-W-W-H-H-H

\bigskip

7F. ~~$(\alpha^2, \alpha^4, \alpha^4, \alpha^4)$

\bigskip

$^*$ have properties $1), ~~2) ~ {\rm and} ~~3)$.

\newpage

\begin{figure}
7B ~~ 4 = (1, 1, 2, 3) ~ 6 = (2, 3, 3, -3, -3, -2)
\begin{center}
\epsfxsize=2.0in
\ \epsfbox{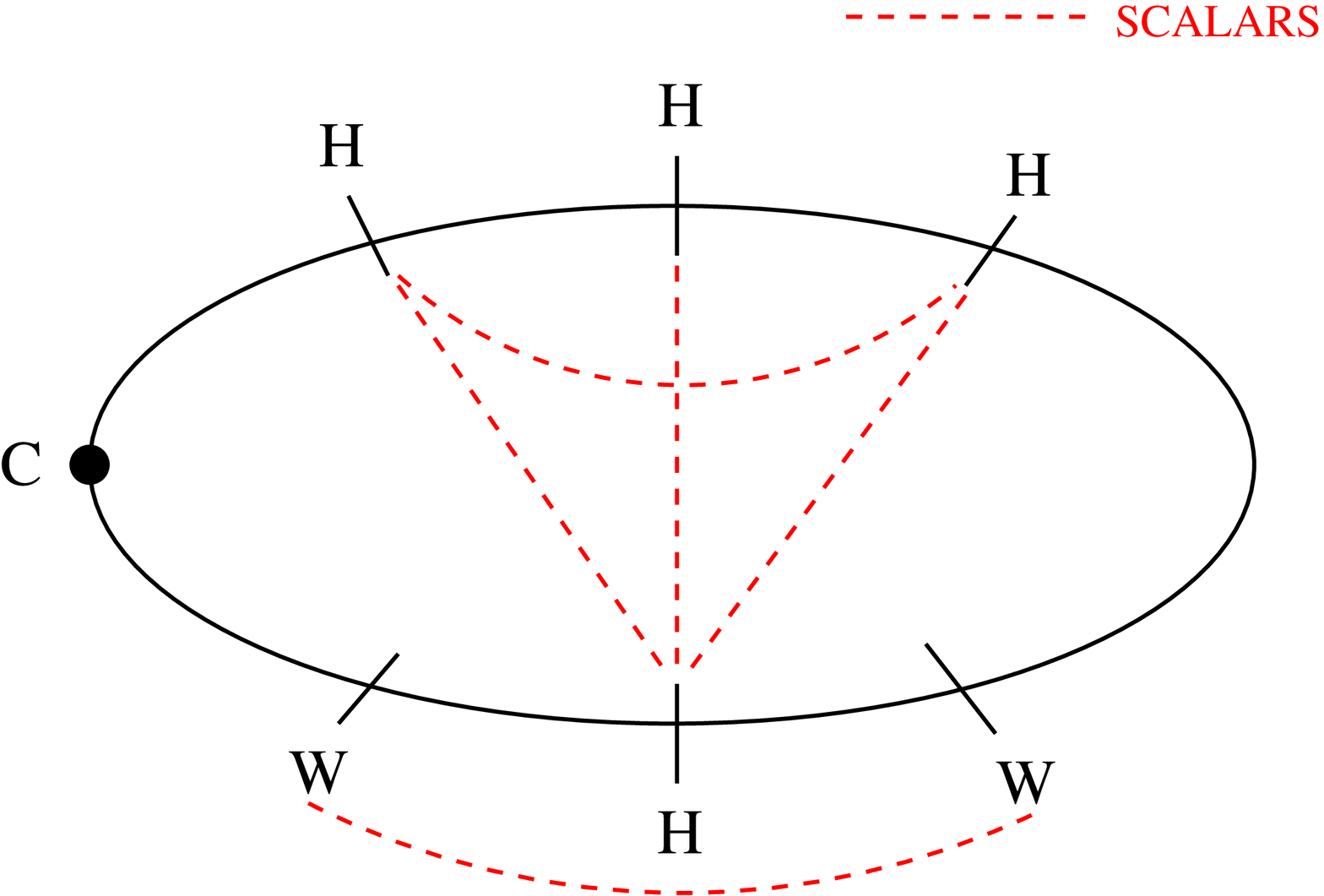}
\end{center}
\end{figure}

~

\begin{figure}
7D ~~ 4 = (1, 3, 5, 5) ~ 6 = (1, 1, 3, -3, -1, -1)
\begin{center}
\epsfxsize=2.0in
\ \epsfbox{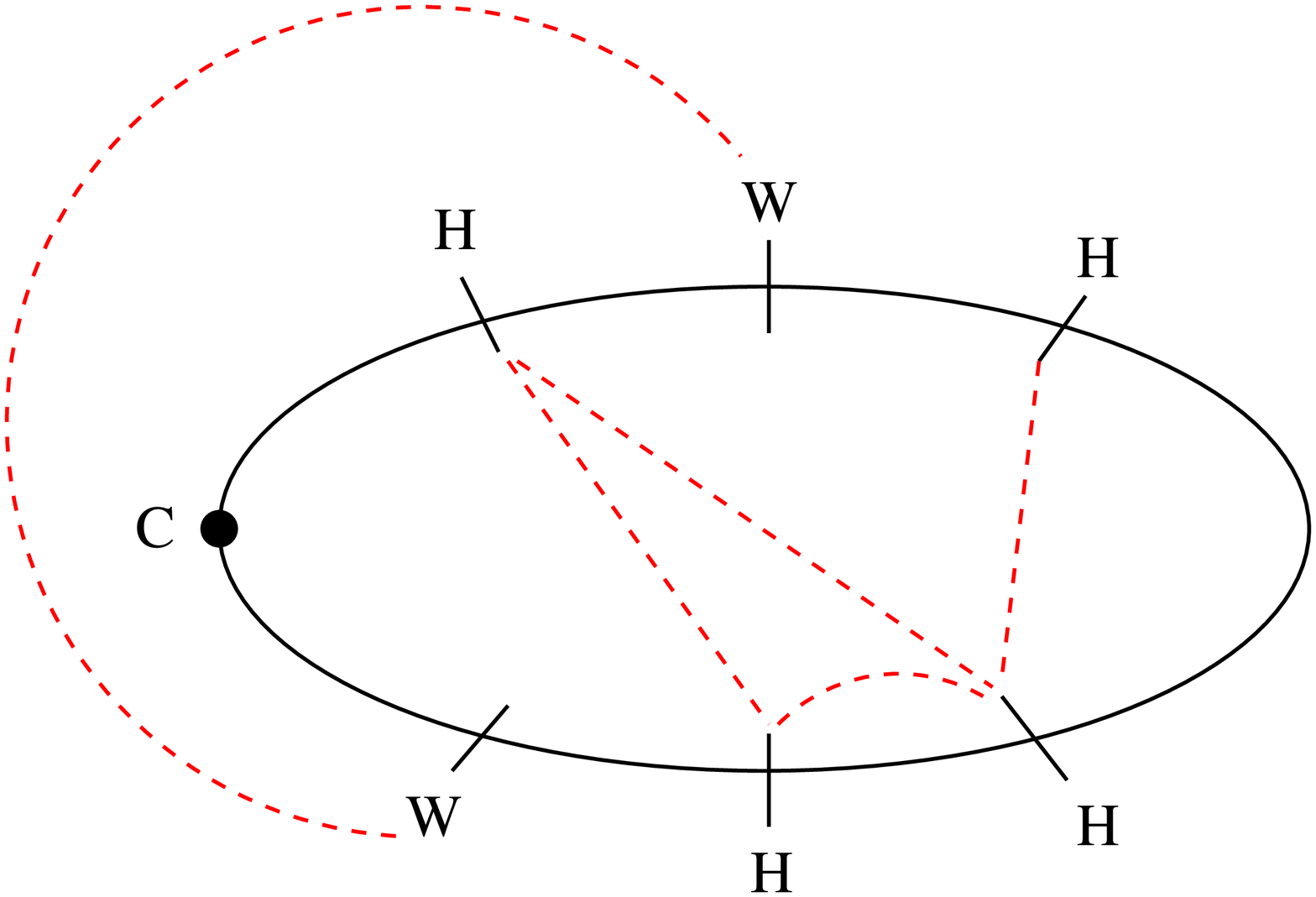}
\end{center}
\end{figure}
\begin{figure}
7E ~~ 4 = (1, 4, 4, 5) ~ 6 = (1, 2, 2, -2, -2, -1)
\begin{center}
\epsfxsize=2.0in
\ \epsfbox{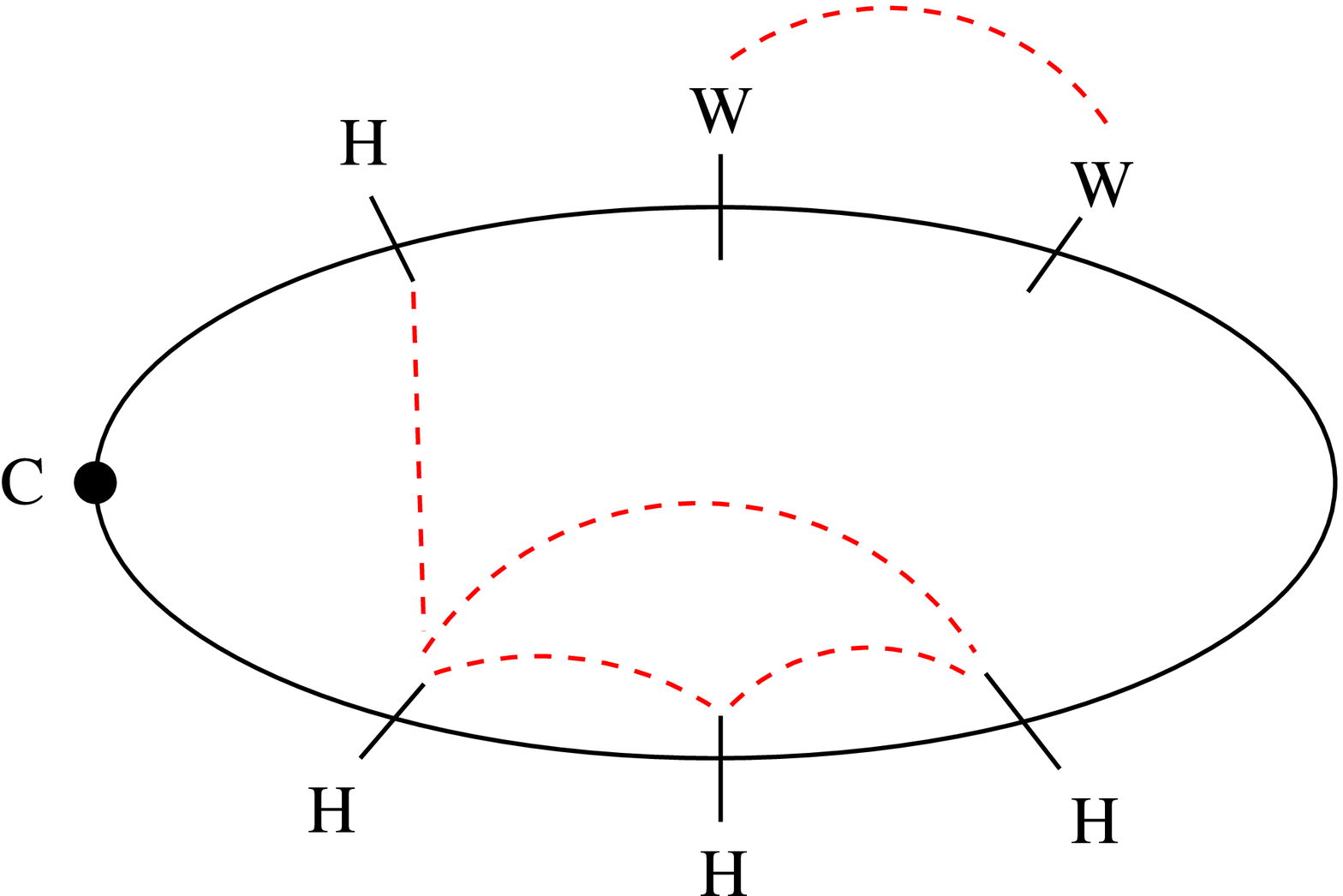}
\end{center}
\end{figure}

~

\newpage

The simplest abelian orbifold conformal extension of the standard model has
$U(3)^7 \rightarrow SU(3)^3$ trinification $\rightarrow (321)_{SM}$.

\bigskip

In this case we have $\alpha_2$ and $\alpha_1$ related correctly for 
low energy. But $\alpha_3(M) \simeq 0.07$ suggesting a conformal scale $M \geq 10$ TeV
 - too high for the L.H.C.

\bigskip
\bigskip

\bigskip
\bigskip

A more unified model which introduces the 4 TeV scale is:

\bigskip
\bigskip

\noindent Taking as orbifold $S^5/Z_{12}$ with embedding of $Z_{12}$
in the SU(4) R-parity  specified by
{\bf 4} $\equiv \alpha^{A_1}, \alpha^{A_2}, \alpha^{A_3}, \alpha^{A_4})$
and $A_{\mu} = (1, 2, 3, 6)$.

\bigskip

\noindent This accommodates the scalars necessary to spontaneously
break to the SM.

\bigskip

\noindent As a bonus, the dodecagonal quiver predicts three chiral families
(see next page). 

\newpage

\begin{figure}
\begin{center}
\epsfxsize=4.0in
\ \epsfbox{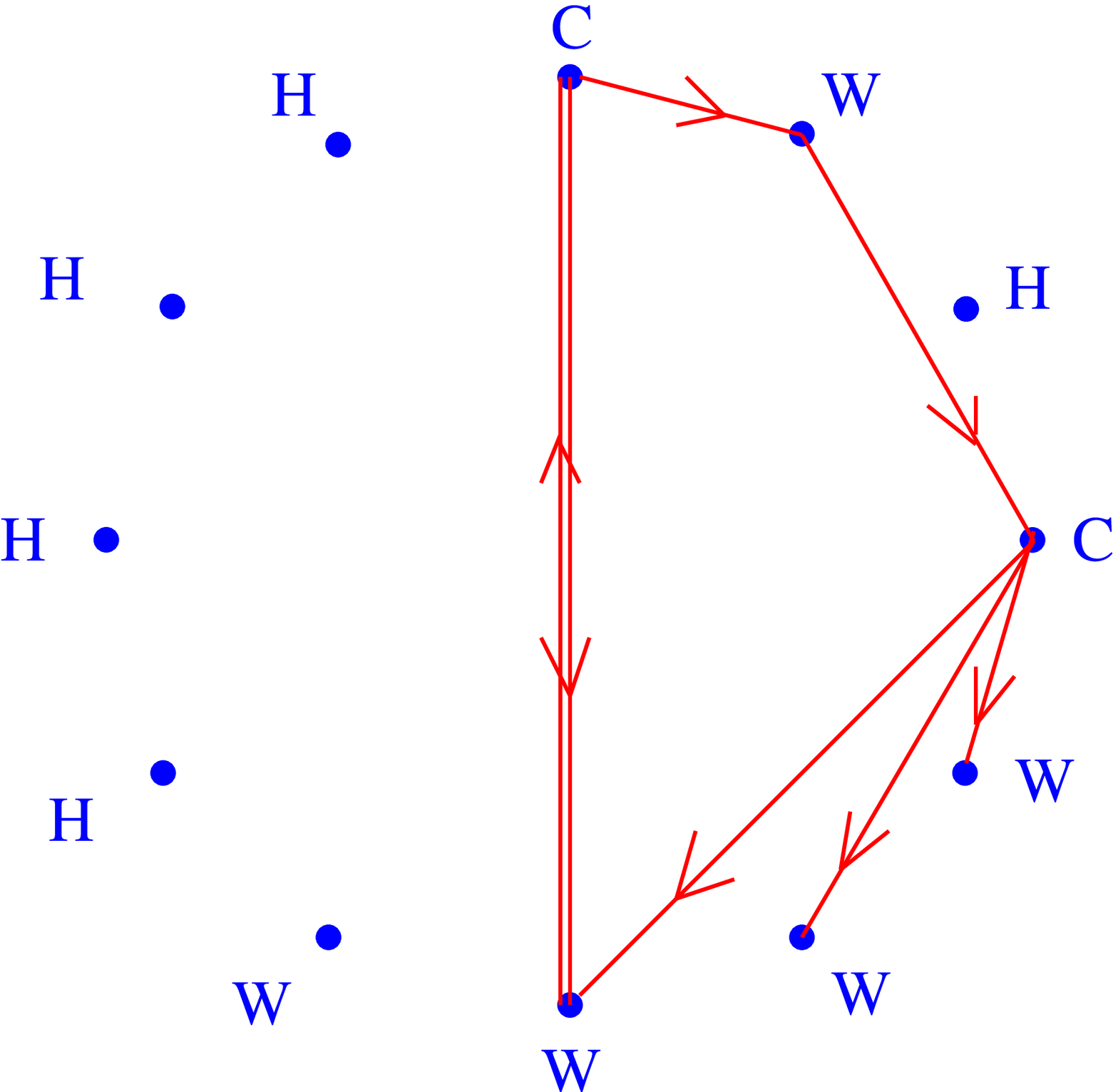}
\end{center}
\end{figure}

\bigskip

\noindent $A_{\mu} = (1, 2, 3, 6)$

\bigskip

\noindent $SU(3)_C \times SU(3)_H \times SU(3)_H$

\bigskip

\noindent $5(3, \bar{3}, 1) + 2 (\bar{3}, 3, 1)$

\newpage

\begin{figure}
\begin{center}
\epsfxsize=6.0in
\ \epsfbox{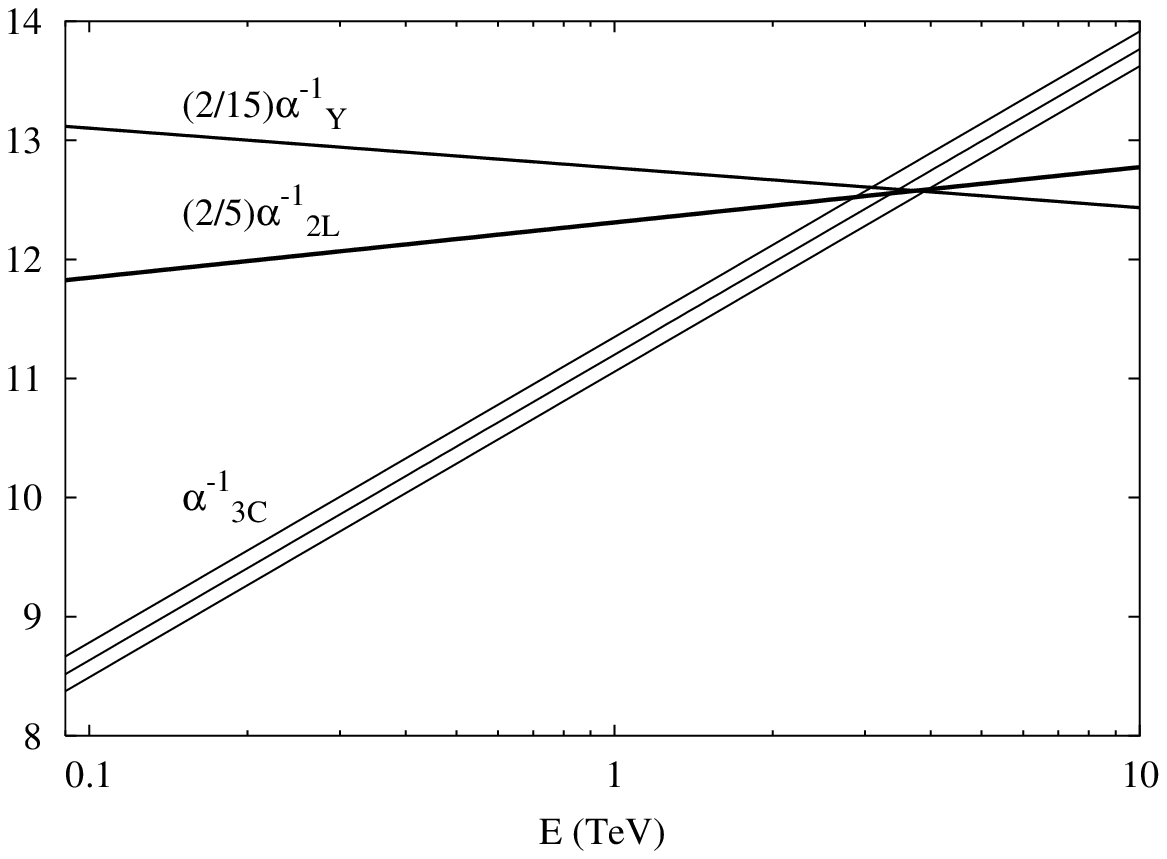}
\end{center}
\end{figure}

\bigskip

\noindent without thresholds: all states at $M_U$

\noindent {\tt hep-ph/0302074}

\noindent PHF+Ryan Rohm + Tomo Takahashi

\newpage

\begin{figure}
\begin{center}
\epsfxsize=6.0in
\ \epsfbox{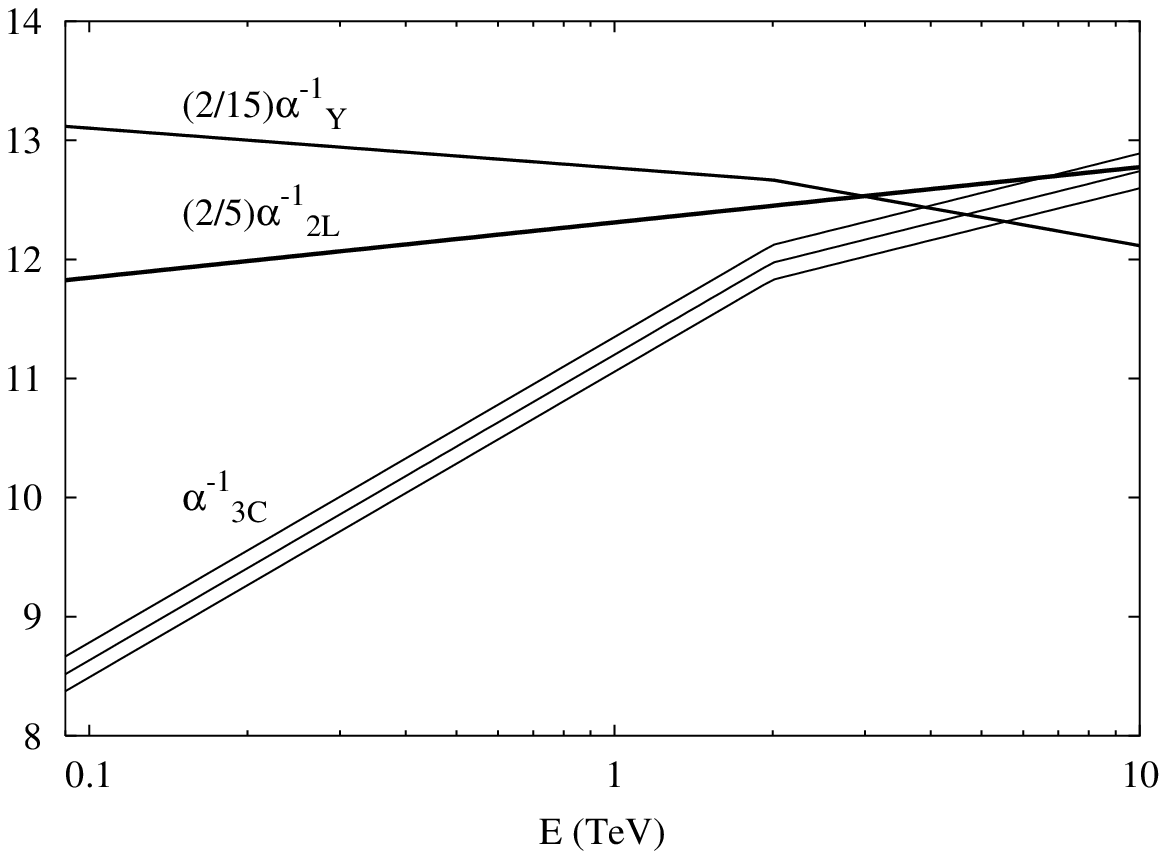}
\end{center}
\end{figure}

\bigskip

\noindent thresholds: CH fermions at 2 TeV

\noindent {\tt hep-ph/0302074}

\noindent PHF+Ryan Rohm + Tomo Takahashi

\newpage

\begin{figure}
\begin{center}
\epsfxsize=6.0in
\ \epsfbox{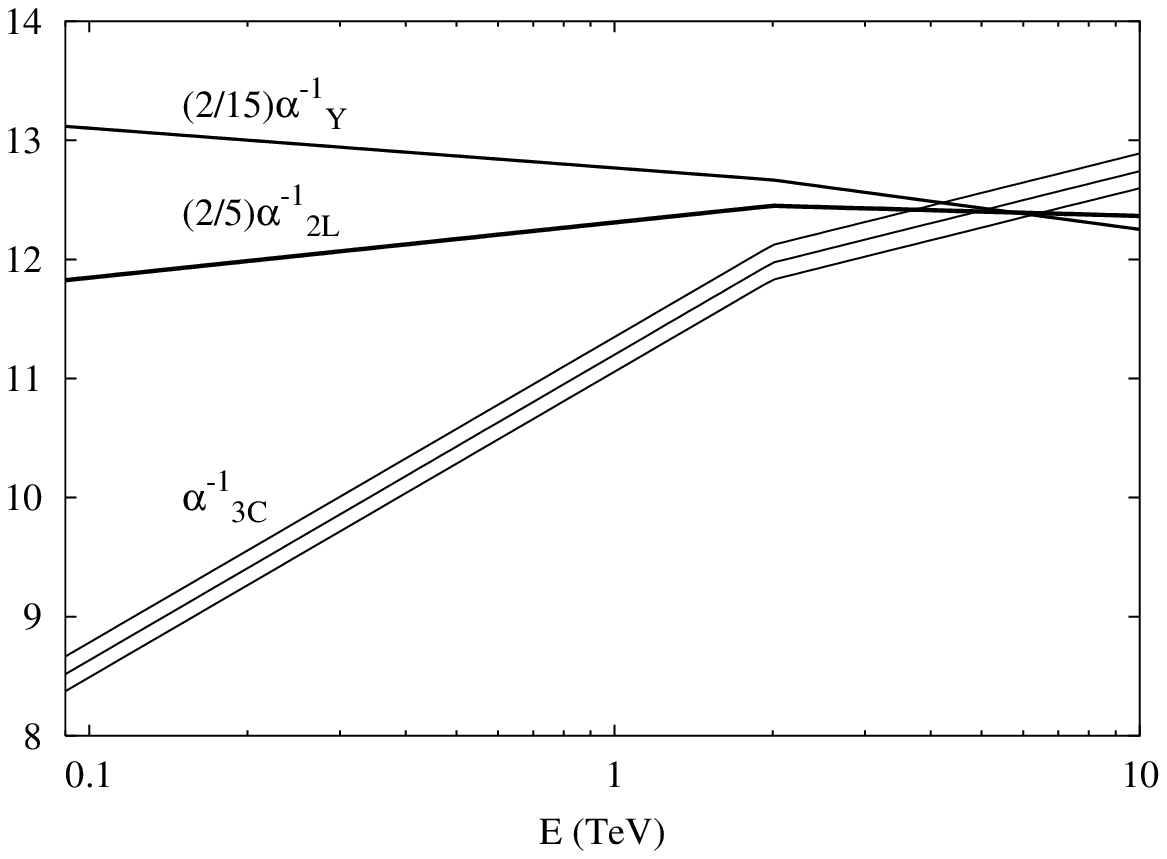}
\end{center}
\end{figure}

\bigskip

\noindent thresholds: CW fermions at 2 TeV

\noindent {\tt hep-ph/0302074}

\noindent PHF+Ryan Rohm + Tomo Takahashi

\newpage

\noindent SECTION 4. QUADRATIC DIVERGENCES.

\bigskip
\bigskip

\noindent \underline{Classification of abelian quiver gauge theories}

\bigskip

We consider the compactification of the type-IIB superstring
on the orbifold $AdS_5 \times S^5/\Gamma$
where $\Gamma$ is an abelian group $\Gamma = Z_p$
of order $p$ with elements ${\rm exp} \left( 2 \pi i A/p \right)$,
$0 \le A \le (p-1)$.

The resultant quiver gauge theory has ${\cal N}$
residual supersymmetries with ${\cal N} = 2,1,0$ depending
on the details of the embedding of $\Gamma$
in the $SU(4)$ group which is the isotropy
of the $S^5$. This embedding is specified
by the four integers $A_m, 1 \le m \le 4$ with

\begin{equation}
\Sigma_m A_m = 0 {\rm (mod p)}
\nonumber
\label{SU4}
\end{equation}

\noindent which characterize
the transformation of the components of the defining
representation of $SU(4)$. We are here interested in the non-supersymmetric
case ${\cal N} = 0$ which occurs if and only if
all four $A_m$ are non-vanishing.

\bigskip
\bigskip

\begin{center}

\bigskip

-----------------------

\bigskip

\end{center}

It can be shown by group theory arguments that:

\bigskip

{\it In an ${\cal N}=0$ quiver gauge theory, }

{\it chiral fermions are present}

{\it if and only if all scalars}

{\it are in bifundamental representations.}

\begin{center}

\bigskip

-----------------------

\bigskip

\end{center}

The following Table lists all abelian quiver
theories with ${\cal N}=0$ from $Z_2$ to $Z_7$:

\bigskip

\newpage

\begin{tabular}{|||c||c||c|c||c|c|c||c|c|||}
\hline
\hline
\hline
& p & $A_m$ & $a_i$ & scal & scal & chir &  \\
&&&& bfds & adjs & frms & SM  \\
\hline
\hline
1 & 2 & (1111) & (000) & 0 & 6 & No  &  No \\
\hline
\hline
2 & 3 & (1122) & (001) & 2 & 4 & No  &  No \\
\hline
\hline
3 & 4 & (2222) & (000) & 0 & 6 & No  &  No \\
4 & 4 & (1133) & (002) & 2 & 4 & No  &  No \\
5 & 4 & (1223) & (011) & 4 & 2 & No  &  No \\
6 & 4 & (1111) & (222) & 6 & 0 & Yes  &  No \\
\hline
\hline
7 & 5 & (1144) & (002) & 2 & 4 & No  &  No \\
8 & 5 & (2233) & (001) & 2 & 4 & No  &  No \\
9 & 5 & (1234) & (012) & 4 & 2 & No  &  No \\
10 & 5 & (1112) & (222) & 6 & 0 & Yes  &  No \\
11 & 5 & (2224) & (111) & 6 & 0 & Yes  &  No \\
\hline
\hline
12 & 6 & (3333) & (000) & 0 & 6 & No  &  No \\
13 & 6 & (2244) & (002) & 2 & 4 & No  &  No \\
14 & 6 & (1155) & (002) & 2 & 4 & No  &  No \\
15 & 6 & (1245) & (013) & 4 & 2 & No  &  No \\
16 & 6 & (2334) & (011) & 4 & 2 & No  &  No \\
17 & 6 & (1113) & (222) & 6 & 0 & Yes  &  No \\
18 & 6 & (2235) & (112) & 6 & 0 & Yes  &  No \\
19 & 6 & (1122) & (233) & 6 & 0 & Yes  &  No \\
\hline
\hline
20 & 7 & (1166) & (002) & 2 & 4 & No  &  No \\
21 & 7 & (3344) & (001) & 2 & 4 & No  &  No \\
22 & 7 & (1256) & (013) & 4 & 2 & No  &  No \\
23 & 7 & (1346) & (023) & 4 & 2 & No  &  No \\
24 & 7 & (1355) & (113) & 6 & 0 & No  &  No \\
25 & 7 & (1114) & (222) & 6 & 0 & Yes  &  No \\
26 & 7 & (1222) & (333) & 6 & 0 & Yes  &  No \\
27 & 7 & (2444) & (111) & 6 & 0 & Yes  &  No \\
28 & 7 & (1123) & (223) & 6 & 0 & Yes  &  Yes \\
29 & 7 & (1355) & (113) & 6 & 0 & Yes  &  Yes \\
30 & 7 & (1445) & (113) & 6 & 0 & Yes  &  Yes \\
\hline
\hline
\hline
\end{tabular}

\bigskip
\bigskip
\bigskip

\newpage

\newpage

\bigskip

{\it Cancellation of quadratic divergences}

The lagrangian for the nonsupersymmetric $Z_p$ theory can be written in
a convenient
notation which accommodates simultaneously both adjoint and
bifundamental scalars as

\small

\begin{eqnarray}
{\cal L} & = &
-\frac{1}{4} F_{\mu\nu; r,r}^{ab}F_{\mu\nu; r,r}^{ba}
+i \bar{\lambda}_{r + A_4, r}^{ab} \gamma^{\mu} D_{\mu} \lambda_{r,
r+A_4}^{ba}
+ 2 D_{\mu} \Phi_{r+a_i, r}^{ab \dagger} D_{\mu} \Phi_{r, r+a_i}^{ba}
+i \bar{\Psi}_{r+A_m, r}^{ab} \gamma^{\mu} D_{\mu} \Psi_{r, r+A_m}^{ba}
\nonumber \\
&  &
- 2 i g
\left[ \bar{\Psi}_{r, r+A_i}^{ab} P_L \lambda_{r + A_i, r + A_i +
A_4}^{bc}
\Phi_{r + A_i+A_4, r}^{\dagger ca}
- \bar{\Psi}_{r, r+A_i}^{ab} P_L \Phi_{r + A_i, r - A_4}^{\dagger bc}
\lambda_{r - A_4, r}^{ca}
\right] \nonumber \\
& & -   \sqrt{2} i g \epsilon_{ijk}
\left[
\bar{\Psi}_{r, r + A_i}^{ab} P_L \Psi_{r +A_i, r + A_i + A_j}^{bc}
\Phi_{r -A_k - A_4, r}^{ca}
-
\bar{\Psi}_{r, r + A_i}^{ab} P_L \Phi_{r +A_i, r + A_i + A_k +
A_4}^{bc} \Psi_{r - A_j, r}^{ca}
\right] \nonumber \\
& & - g^2 \left(
\Phi_{r, r + a_i}^{ab} \Phi_{r+a_i,r}^{\dagger bc}
-
\Phi_{r, r - a_i}^{\dagger ab} \Phi_{r-a_i,r}^{bc}
\right)
\left(
\Phi_{r, r + a_j}^{cd} \Phi_{r+a_j,r}^{\dagger da}
-
\Phi_{r, r - a_j}^{\dagger cd} \Phi_{r- a_j,r}^{da}
\right)  \nonumber \\
& & + 4 g^2
\left(
\Phi_{r, r+a_i}^{ab}\Phi_{r+a_i, r+a_i+a_j}^{bc}
\Phi_{r+a_i+a_j,r+a_j}^{\dagger cd}\Phi_{r+a_j,r}^{\dagger da}
-
\Phi_{r, r+a_i}^{ab}\Phi_{r+a_i, r+a_i+a_j}^{bc}
\Phi_{r+a_i+a_j,r+a_i}^{\dagger cd}\Phi_{r+a_i, r}^{\dagger da}
\right)
\nonumber
\label{N=0L}
\end{eqnarray}

\large

\noindent where $\mu, \nu = 0, 1, 2, 3$ are lorentz indices; $a, b, c, d = 1$ to
$N$ are $U(N)^p$
group labels; $r = 1$ to $p$ labels the node of the quiver diagram
;$a_i ~~ (i = \{1, 2, 3\}) $ label the first three of the {\bf 6} of
SU(4);
$A_m ~~ (m = \{1, 2, 3 ,4\}) = (A_i, A_4)$ label the {\bf 4} of SU(4).
By definition
$A_4$ denotes an arbitrarily-chosen fermion ($\lambda$) associated with
the gauge boson,
similarly to the notation in the ${\cal N} = 1$ supersymmetric case.

\bigskip
\bigskip
\bigskip

Let us first consider the quadratic divergence question in the mother
${\cal N} = 4$ theory. The ${\cal N}=4$ lagrangian is like
Eq.(\ref{N=0L})
but since there is only one node all those subscripts become
unnecessary
so the form is simply

\begin{eqnarray}
{\cal L} & = &
-\frac{1}{4} F_{\mu\nu}^{ab}F_{\mu\nu}^{ba}
+i \bar{\lambda}^{ab} \gamma^{\mu} D_{\mu} \lambda^{ba}
+ 2 D_{\mu} \Phi_{i}^{ab \dagger} D_{\mu} \Phi_{i}^{ba}
+i \bar{\Psi}_{m}^{ab} \gamma^{\mu} D_{\mu} \Psi_{m}^{ba} \nonumber \\
&  &
- 2 i g
\left[\bar{\Psi}_{i}^{ab} P_L \lambda^{bc}
\Phi_{i, r}^{\dagger ca}
- \bar{\Psi}_{i}^{ab} P_L \Phi_{i}^{bc}\lambda^{ca}
\right] \nonumber \\
& & - \sqrt{2} i g \epsilon_{ijk}
\left[
\bar{\Psi}_{i}^{ab} P_L \Psi_{j}^{bc} \Phi_{k}^{\dagger ca}
-
\bar{\Psi}_{i}^{ab} P_L \Phi_{j}^{bc} \Psi_{k}^{ca}
\right] \nonumber \\
& & - g^2 \left(
\Phi_{i}^{ab} \Phi_{i}^{\dagger bc}
-
\Phi_{i}^{\dagger ab} \Phi_{i}^{bc}
\right)
\left(
\Phi_{j}^{cd} \Phi_{j}^{\dagger da}
-
\Phi_{j}^{\dagger cd} \Phi_{j}^{da}
\right)  \nonumber \\
& & + 4 g^2
\left(
\Phi_{i}^{ab}\Phi_{j}^{bc}
\Phi_{i}^{\dagger cd}\Phi_{j}^{\dagger da}
-
\Phi_{i}^{ab}\Phi_{j}^{bc}
\Phi_{j}^{\dagger cd}\Phi_{i}^{\dagger da}
\right)
\nonumber
\label{N=4L}
\end{eqnarray}

\newpage

\bigskip

All ${\cal N} = 4$ scalars are in adjoints and the scalar propagator
has one-loop quadratic divergences coming potentially from
three scalar self-energy diagrams:
(a) the gauge loop (one quartic vertex);
(b) the fermion loop (two trilinear vertices);
and (c) the scalar loop (one quartic vertex).

For ${\cal N} = 4$ the respective contributions
of (a, b, c) are computable from
Eq.(\ref{N=4L}) as proportional to $g^2N (1, -4, 3)$ which cancel
exactly.

\bigskip
\bigskip
\bigskip

The ${\cal N} = 0$ results for the scalar self-energies (a, b, c)
are computable from the lagrangian of Eq.(\ref{N=0L}).
Fortunately, the calculation was already done in the literature.
The result is amazing! The quadratic divergences cancel
if and only if x = 3, exactly the same ``if and only if"
as to have chiral fermions.
It is pleasing that one can independently confirm
the results directly from the interactions
in Eq.(\ref{N=0L}) To give just one explicit
example, in the contributions to
diagram (c) from the last term in Eq.(\ref{N=0L}), the
1/N corrections arise from a contraction of $\Phi$ with
$\Phi^{\dagger}$
when all the four color superscripts are distinct
and there is consequently no sum
over color in the loop. For this case, examination of the node
subscripts then
confirms proportionality to the kronecker delta, $\delta_{0, a_i}$.
If and only if all $a_i \neq 0$, all the other terms
in Eq.(\ref{N=0L}) do not lead to 1/N corrections
to the ${\cal N}=4$.

\newpage

\bigskip

\noindent SECTION 5. ANOMALY CANCELLATION AND CONFORMAL U(1)s  

\bigskip
\bigskip

The lagrangian for the nonsupersymmetric $Z_n$ theory can be written in
a convenient
notation which accommodates simultaneously both adjoint and
bifundamental scalars as mentioned before.

As also mentioned above we shall restrict attention to models
where all scalars are in bifundamentals which
requires all $a_i$ to be non zero. Recall that
$a_1=A_2+A_3$, $a_2=A_3+A_1$; $a_3=A_1+A_2$.

\bigskip
\bigskip
\bigskip

The lagrangian is classically $U(N)^p$ gauge
invariant. There are, however, triangle anomalies of the
$U(1)_p U(1)^2_q$ and $U(1)_p SU(N)_q^2$ types. Making gauge transformations under
the $U(1)_r$ (r = 1,2,...,n) with gauge parameters $\Lambda_r$
leads to a variation

\begin{equation}
\delta {\cal L} = - \frac{g^2}{4\pi^2}\Sigma_{p=1}^{p=n} A_{pq} F_{\mu\nu}^{(p)}
\tilde{F}^{(p) \mu\nu} \Lambda_q
\label{Apq}
\end{equation}
which defines an $n \times n$ matrix $A_{pq}$ which is given by

\begin{equation}
A_{pq} = {\rm Tr} (Q_p Q_q^2)
\label{chiraltrace}
\end{equation}
where the trace is over all chiral fermion links and $Q_r$ is the
charge of the bifundamental under $U(1)_r$. We shall adopt the sign
convention that ${\bf N}$ has $Q=+1$ and ${\bf N^{*}}$ has $Q=-1$.

\bigskip
\bigskip
\bigskip

It is straightforward to write $A_{pq}$ in terms of
Kronecker deltas because the content of chiral fermions is

\begin{equation}
\Sigma_{m=1}^{m=4} \Sigma_{r=1}^{r=n} ({\bf N}_r, {\bf N^{*}}_{r+A_{m}})
\end{equation}
This implies that the antisymmetric matrix $A_{pq}$ is explicitly

\begin{equation}
A_{pq} = - A_{qp} = \Sigma_{m=1}^{m=4} \left( \delta_{p, q-A_{m}} -
\delta_{p, q+A_{m}} \right)
\label{ApqKronecker}
\end{equation}

\bigskip

\newpage

\bigskip

Now we are ready to construct ${\cal L}_{comp}^{(1)}$, the compensatory
term. Under the $U(1)_r$ gauge transformations with
gauge parameters $\Lambda_r$ we require that

\begin{eqnarray}
\delta {\cal L}_{comp}^{(1)} & = &   - \delta {\cal L} \nonumber \\
& = & + \frac{g^2}{4\pi^2}\Sigma_{p=1}^{p=n} A_{pq} F_{\mu\nu}^{(p)}
\tilde{F}^{(p) \mu\nu} \Lambda_q
\label{compensatory}
\end{eqnarray}
To accomplish this property, we 
construct a compensatory term in the form
\begin{equation}
{\cal L}_{comp}^{(1)} = \frac{g^2}{4 \pi} \Sigma_{p=1}^{p=n}
\Sigma_{k} B_{pk} {\rm Im} {\rm Tr} {\rm ln}
\left( \frac{\Phi_k}{v} \right) F_{\mu\nu}^{(p)} \tilde{F}^{(p) \mu\nu}
\label{compensatory2}
\end{equation}
where $\Sigma_{k}$ runs over scalar links.
To see
that ${\cal L}_{comr}^{(1)}$ of Eq.(\ref{compensatory2}) has
$SU(N)^n$.
invariance
rewrite Tr ln $\equiv$ exp det and note that the $SU(N)$
matrices have unit determinant.

\bigskip
\bigskip
\bigskip

We note {\it en passant} that one cannot take the
$v \rightarrow 0$ limit in Eq.(\ref{compensatory2}); the chiral anomaly
enforces a breaking of conformal invariance.

We define a matrix $C_{kq}$ by

\begin{equation}
\delta \left( \Sigma_{p=1}^{p=n} \Sigma_k {\rm Im}
{\rm Tr} {\rm ln} \left( \frac{\Phi_k}{v} \right) \right)
= \Sigma_{q=1}^{q=n} C_{kq} \Lambda_q
\label{Ckq}
\end{equation}
whereupon Eq.(\ref{compensatory}) will be satisfied if
the matrix $B_{pk}$ satisfies $A=BC$. The inversion $B=AC^{-1}$ is
non trivial because $C$ is singular but $C_{kq}$ can be written in terms
of Kronecker deltas by noting that the content of complex scalar fields in the model
implies that the matrix $C_{kq}$ must be of the form
\begin{equation}
C_{kq} = 3 \delta_{kq} - \Sigma_{i} \delta_{k+a_{i},q}
\label{CkqKronecker}
\end{equation}

\bigskip

\newpage

\bigskip

{\it Evolution of U(1) gauge couplings.}

\bigskip

In the absence of the compensatory term, the two independent $U(N)^n$
gauge couplings $g_N$ for SU(N) and $g_1$ for U(1) are taken to be
equal $g_N(\mu_0) = g_1(\mu_0)$ at a chosen scale, {\it e.g.} $\mu_0$=4 TeV
to enable cancellation of quadratic
divergences. Note that the $n$ SU(N) couplings
$g_N^{(p)}$ are equal by the overall $Z_n$ symmetry,
as are the $n$ U(1) couplings $g_1^{(p)}$, $1 \le p \le n$.

As one evolves to higher scales $\mu > \mu_0$, the renormalization
group beta function $\beta_N$ for SU(N)
vanishes $\beta_N =0$ at least at one-loop level so
the $g_N(\mu)$ can behave independent of the scale as expected by conformality.
On the other hand, the beta function $\beta_1$ for
U(1)
is positive definite in the unadorned theory, given at one loop by
\begin{equation}
b_1 = \frac{11N}{48\pi^2}
\label{b1}
\end{equation}
where N is the number of colors.

\bigskip
\bigskip
\bigskip

The corresponding coupling satisfies
\begin{equation}
\frac{1}{\alpha_1(\mu)} = \frac{1}{\alpha_1(M)} + 8\pi b_1 {\rm ln} \left( \frac{M}{\mu}
\right)
\end{equation}
so the Landau pole, putting $\alpha(\mu)=0.1$ and $N=3$, occurs at
\begin{equation}
\frac{M}{\mu} = {\rm exp} \left[ \frac{20 \pi}{11} \right] \simeq 302
\end{equation}
so for $\mu = 4$ TeV, $M \sim 1200$ TeV. The coupling becomes ``strong''
$\alpha(\mu) = 1$ at
\begin{equation}
\frac{M}{\mu} = {\rm exp} \left[ \frac{18 \pi}{11} \right] \simeq 171
\end{equation}
or $M \sim 680$ TeV.

We may therefore ask whether the new term ${\cal L}_{comp}$
in the lagrangian, necessary for anomaly cancellation, can
solve this problem for conformality?

\bigskip

\newpage

\bigskip

Since the scale $v$ breaks conformal invariance,
the matter fields acquire mass, so the one-loop
diagram \footnote{The usual one-loop $\beta-$function is
of order $h^2$ regarded as an expansion in Planck's constant:
four propagators each $\sim h$ and two vertices each $\sim h^{-1}$
(c.f. Y. Nambu, Phys. Lett. {\bf B26,} 626 (1968)).
The diagram considered
is also $\sim h^2$ since it has three propagators,
one quantum vertex $\sim h$ and an additional $h^{-2}$ associated with
$\Delta m^2_{kk'}$.} has a logarithmic divergence proportional to

\begin{equation}
\int \frac{d^4p}{v^2} \left[ \frac{1}{(p^2-m_k^2)} - \frac{1}{(p^2-m_{k'}^2)}
\right]
\sim - \frac{ \Delta m^2_{kk'}}{v^2} {\rm ln} \left( \frac{\Lambda}{v} \right)
\end{equation}
the sign of which depends on $\delta m^2_{kk'} = (m_k^2 - m_{k'}^2)$.

\bigskip
\bigskip
\bigskip

\noindent To achieve conformality of U(1), a constraint must be imposed
on the mass spectrum of matter bifundamentals, {\it viz}
\begin{equation}
\Delta m^2_{kk'} \propto v^2 \left( \frac{11N}{48\pi^2} \right)
\end{equation}
\noindent with a proportionality constant of order one which depends on the choice
of model, the $n$ of $Z_n$ and the values chosen for $A_m, m=1,2,3$. This
signals how conformal invariance must be broken at
the TeV scale in order that it can be restored at high energy;
it is interesting that such a constraint arises
in connection with an anomaly cancellation mechanism
which necessarily breaks conformal symmetry.

\bigskip
\bigskip

\newpage

\bigskip

\noindent SECTION 6. DARK MATTER CANDIDATE

\bigskip
\bigskip

\noindent {\it Definition of a $Z_2$ symmetry}

\bigskip

In the nonsupersymmetric quiver gauge theories, the
gauge group, for abelian orbifold $AdS_5 \times S^5/Z_n$
is $U(N)^n$. In phenomenological application $N=3$
and $n$ reduces eventually after symmetry breaking
to $n=3$ as in trinification.
The chiral fermions are then in the representation of $SU(3)^3$:
\begin{equation}
(3, 3^*, 1) + (3^*, 1, 3) + (1, 3, 3^*)
\end{equation}
This is as in the {\bf 27} of $E_6$ where the particles
break down in to the following representations of the
$SU(3) \times SU(2) \times U(1)$ standard model group:
\begin{equation}
Q, ~~~~~ u^c, ~~~~ d^c, ~~~~ L ~~~~ e^c ~~~~~ N^c
\end{equation}
transforming as
\begin{equation}
(3,2), ~~ (3^*,1), ~~ (3^*,1), ~~ (1,2), ~~ (1,1), ~~ (1,1) \\
\end{equation}
in a {\bf 16} of the $SO(10)$ subgroup.

\bigskip
\bigskip

\noindent In addition there are the states
\begin{equation}
h, ~~~~~ h*, ~~~~~ E, ~~~~~ E*
\end{equation}
transforming as
\begin{equation}
(3, 1), ~~ (3^*, 1), ~~ (2,1), ~~ (2,1)
\end{equation}
in a {\bf 10} of $SO(10)$
and finally
\begin{equation}
S
\end{equation}
transforming as the singlet
\begin{equation}
(1, 1)
\end{equation}

It is natural to define a $Z_2$ symmetry $R$ which commutes
with the $SO(10)$ subgroup of $E_6 \rightarrow O(10) \times U(1)$
such that $R=+1$ for the first {\bf 16} of states.
Then it is mandated that $R=-1$ for the {\bf 10} and {\bf 1}
of SO(10) because the following Yukawa couplings must
be present to generate mass for the fermions:
\begin{equation}
16_f 16_f 10_s, ~~~~16_f 10_f 16_s, ~~~~ 10_f 10_f 1_s, ~~~~
10_f 1_f 1_s
\end{equation}
\noindent which require $R=+1$ for $10_s, 1_s$ and $R=-1$ for $16_s$.

\newpage

\bigskip

\noindent Contribution of the Lightest Conformality Particle (LCP) to the cosmological energy density:

\bigskip

The LCP act as cold dark matter WIMPs, and the calculation of the
resultant energy density follows a well-known path. Here
we follow the procedure in a recent technical book by Mukhanov.

The LCP decouple at temperature $T_*$, considerably less than their
mass $M_{LCP}$; we define $x_* = M_{LCP}/T_*$. Let the
annihilation cross-section of the LCP
at decoupling be $\sigma_*$. Then the dark matter density
$\Omega_m$, relative to the critical density,
\begin{equation}
\Omega_m h_{75}^2 = \frac{\tilde{g}_*^{1/2}}{g_*} x_*^{3/2} \left(
\frac{3 \times 10^{-38} cm^2}{\sigma_*} \right)
\label{OmegaM}
\end{equation}
where $h_{75}$ is the Hubble constant in units
of $75km/s/Mpc$.
$g_* = (g_b + \frac{7}{8}g_f)$ is the effective number of
degrees of freedom (dof) at freeze-out for all particles
which later convert their energy into photons;
and $\tilde{g}_*$ is the number
of dof which are relativistic at $T_*$.

\bigskip
\bigskip
\bigskip

\noindent Discussion of LCP Dark Matter:

\bigskip

\noindent The LCP is a viable candidate for a
cold dark matter particle which can be produced at the LHC.
The distinction from other candidates
will require establishment of the $U(3)^3$ gauge bosons, extending the 3-2-1 standard model
and the discovery that the LCP is in a bifundamental
representation thereof.

\noindent To confirm that the LCP is the dark matter particle would, however,
require direct detection of dark matter.

\newpage

{\it Status of conformality}

\bigskip
\bigskip

\noindent It has been established that
conformality
can provide (i) naturalness without one-loop
quadratic divergence for the scalar mass and
anomaly cancellation;
(ii) precise unification of the coupling constants;
and (iii) a viable dark matter candidate.
It remains for experiment to check that quiver gauge
theories with gauge group $U(3)^3$ or $U(3)^n$ with $n \geq 4$
are actually employed by Nature.

\bigskip
\bigskip
\bigskip

\noindent SECTION 7. SUMMARY

\bigskip
\bigskip

\begin{itemize}

\item Phenomenology of conformality has striking resonances
with the standard model.

\item 4 TeV Unification predicts three families and new particles
around 4 TeV accessible to experiment (LHC).

\item The scalar propagator in these theories has no
quadratic divergence iff there are chiral fermions.

\item Anomaly cancellation in effective lagrangian
connected to consistency of U(1) factors.

\item Dark matter candidate (LCP) will be
produced at LHC, then directly detected.

\end{itemize}

\bigskip

\begin{center}

{\bf Acknowledgements}

\end{center}

\bigskip

This work was supported in part by
the U.S. Department of Energy under Grant No. DE-FG02-97ER-41036.

\newpage

\noindent BIBLIOGRAPHY FOR CONFORMALITY

\bigskip

This is an attempt at a  complete bibliography for conformality.
I found only 23 papers
and will be grateful if anyone can notify me of omissions.   

\bigskip

\noindent The first paper with the general idea was

P.H. Frampton, Phys.Rev. {\bf D60,} 041901 (1999). {\tt hep-th/9812117}.

\noindent Shortly therafter was a paper with a student:

P.H. Frampton and W.F. Shively, Phys. Lett. {\bf B454,} 49 (1999). {\tt hep-th/9902168}.

\noindent{\it Caution}: The group theory herein was misunderstood and hence not consistent; 
see later papers for correct and consistent model-building.

\noindent Aspects of conformality phenomenology were given in

P.H. Frampton and C. Vafa. {\tt hep-th/9903226}.

\noindent This paper was not submitted for publication.

\noindent Model building for abelian orbifolds is in

P.H. Frampton, Phys. Rev. {\bf D60,} 087901 (1999). {\tt hep-th/9905042}.

P.H. Frampton, Phys. Rev. {\bf D60,} 121901 (1999). {\tt hep-th/9907051}.

\noindent Model building based on non-abelian orbifolds is in

P.H. Frampton and T.W. Kephart, Phys. Lett. {\bf B485,} 403 (2000). {\tt hep-th/9912028}.

P.H. Frampton and T.W. Kephart, Phys. Rev. {\bf D64,} 086007 (2001). {\tt hep-th/0011186}.

P.H. Frampton, R.N. Mohapatra and S. Suh, Phys. Lett. {\bf B520,} 331 (2001).
{\tt hep-ph/0104211}.

\noindent Renormalization group considerations are in

P.H. Frampton and P. Minkowski. {\tt hep-th/0208024}.

\noindent Unification of strong and electroweak forces was in

P.H. Frampton, Mod. Phys. Lett. {\bf A18,} 1377 (2003). {\tt hep-ph/0208044}.

P.H. Frampton, R.M. Rohm and T. Takahashi, Phys. Lett. {\bf B570,} 67 (2003).
{\tt hep-ph/0302074}.

\newpage

\noindent Further discussions of model building are in

T.W. Kephart and H. Paes, Phys. Lett. {\bf B522,} 315 (2001). {\tt hep-ph/0109111}.

P.H. Frampton and T.W. Kephart, Phys. Lett. {\bf B585,} 24 (2004). {\tt hep-th/0306053}.

P.H. Frampton and T.W. Kephart, Int. J. Mod. Phys. {\bf A19,} 593 (2004). {\tt hep-th/0306207}.

T.W. Kephart and H. Paes, Phys. Rev. {\bf D70,} 086009 (2004). {\tt hep-ph/0402228}.

T.W. Kephart, C.A. Lee and Q. Shafi. {\tt hep-ph/0602055}.

\noindent Quadratic divergences and naturalness of the mini-hierarchy are in  

X. Calmet, P.H. Frampton and R.M. Rohm, Phys. Rev. {\bf D72,} 055005 (2005).
{\tt hep-th/0412176}

\noindent The quadratic divergence was calculated in

P. Brax, A. Falkowski, Z. Lalak and S. Pokorski, Phys. Lett. {\bf B538,} 426 (2002). {\tt hep-th/0204195}.

\noindent An earlier correct calculation ignoring U(1)s was 

C. Csaki, W. Skiba and J. Terning, Phys. Rev. {\bf D61,} 025019 (2000).
{\tt hep-th/9906057}.

\noindent The necessity of including U(1)s was in

E. Fuchs, JHEP 0010 (2000) 028. 
{\tt hep-th/0003235}.
 
\noindent A suggestion of a global symmetry was made in 

P.H. Frampton, Mod. Phys. Lett. {\bf A21,} 893 (2006).
{\tt hep-th/0511265}.

\noindent Anomaly cancellation and conformal U(1)s are in

E. Di Napoli and P.H. Frampton, Phys. Lett. {\bf B638,} 374 (2006). {\tt hep-th/0603065}.

\noindent A similar anomaly cancellation in a different context is in

E. Dudas, A. Falkowski and S. Pokorski, Phys. Lett. {\bf B568,} 281 (2003). {\tt hep-th/0303255}.

\noindent Finally, for the Lightest Conformality Particle (LCP) as cold dark matter 

P.H. Frampton. {\tt astro-ph/0607391}.

\end{document}